\documentclass[runningheads]{llncs}
\usepackage{xcolor}
\usepackage{amsmath}
\usepackage{amssymb}

\newif\ifblind
\newif\ifdraft
\blindfalse
\draftfalse
\usepackage{cite}
\usepackage[utf8]{inputenc}
\usepackage{graphicx}
\usepackage{booktabs} 
\usepackage[scaled=0.75]{beramono}
\usepackage[T1]{fontenc}
\usepackage{adjustbox}  
\usepackage{xspace}
\usepackage{wrapfig}
\ifblind
\newcommand{\TRUSTED}{\textsc{<PROJECT>}\xspace}
\else
\newcommand{\TRUSTED}{\textsc{Trusted}\xspace}
\fi

\usepackage{listings}
\definecolor{amber}{rgb}{1.0, 0.75, 0.0}

\ifdraft
\newcommand{\me}[1]{\textcolor{red!40!blue}{[\textbf{ME: }#1]}}
\newcommand{\dn}[1]{\textcolor{blue}{[\textbf{DN: }#1]}}
\newcommand{\sebc}[1]{\textcolor{red}{[\textbf{SC: }#1]}}
\newcommand{\pe}[1]{\textcolor{green!50!black}{[\textbf{PE: }#1]}}
\newcommand{\sm}[1]{\textcolor{amber!50!red}{[\textbf{SM: }#1]}}
\newcommand{\rbnote}[1]{\textcolor{red}{[\textbf{RB: }#1]}}
\else
\newcommand{\me}[1]{}
\newcommand{\dn}[1]{}
\newcommand{\sebc}[1]{}
\newcommand{\pe}[1]{}
\newcommand{\sm}[1]{}
\newcommand{\rbnote}[1]{}
\fi

\usepackage[hyphens]{url}

\usepackage[
pdftitle={A Systematic Approach to Automotive Security},
pdfkeywords={Cybersecurity, Testing, Automotive, Threats},
bookmarks=true,
breaklinks=true,
colorlinks=true,
linkcolor=black,
citecolor=black,
urlcolor=black,
pdfpagelayout=SinglePage,
pdfstartview=Fit
]{hyperref}

\usepackage{multicol}
\usepackage{caption}
\usepackage{subcaption}
\usepackage[usestackEOL]{stackengine}
\setstackgap{S}{0pt}

\usepackage{enumitem}
\usepackage{tikz}
\usetikzlibrary{arrows,automata,positioning,calc,trees,arrows,arrows.meta}

\newcommand*\circled[1]{\tikz[baseline=(char.base)]{
            \node[shape=circle,draw,inner sep=0.35pt] (char) {\small{#1}};}}

\lstdefinelanguage{SMT2}{
    keywords={FLOW, SOURCE, ELEMENT, NOT, IN, HOLDS, ASSET, INCLUDES, HAS, NO, CONNECTOR, TARGET, PROVIDES, CAPABILITY},
    keywords=[2]{state,attr,type},
    keywords=[3]{StateId},
    keywordstyle={\color{blue}\bfseries},
    keywordstyle=[2]{\color{red!50!blue}},
    keywordstyle=[3]{\color{teal}},
    sensitive=false, 
    morecomment=[l]{\#}, 
    showstringspaces=false,
} %

\let\llncssubparagraph\subparagraph
\let\subparagraph\paragraph
\usepackage{titlesec}
\let\subparagraph\llncssubparagraph
\titlespacing*\section{0pt}{10pt plus 2pt minus 2pt}{8pt plus 2pt minus 2pt}
\titlespacing*\subsection{0pt}{10pt plus 2pt minus 2pt}{6pt plus 2pt minus 2pt}

\begin{document}
\title{A Systematic Approach to Automotive Security}

\newcommand{\TUG}{Graz University of Technology}
\newcommand{\AIT}{Austrian Institute of Technology}
\newcommand{\AVL}{AVL List GmbH}
\newcommand{\MDU}{M\"{a}lardalen University}

%
\author{Masoud Ebrahimi\inst{1} \and
Stefan Marksteiner\inst{2,4} \and
Dejan Ničković\inst{3} \and
Roderick Bloem\inst{1} \and
David Schögler\inst{2} \and
Philipp Eisner\inst{2} \and
Samuel Sprung\inst{2} \and 
Thomas Schober\inst{2} \and
Sebastian Chlup\inst{3} \and
Christoph Schmittner\inst{3} \and
Sandra König\inst{3}}
\authorrunning{M. Ebrahimi et al.}
%
\institute{\TUG, Graz, Austria\\
\email{ebrahimi@tugraz.at}\\
\and
\AVL, Graz, Austria \\
\email{stefan.marksteiner@avl.com}\\
\and
\AIT, Vienna, Austria\\
\email{dejan.nickovic@ait.ac.at}
\and
\MDU, V\"{a}ster\r{a}s, Sweden\\
}
\maketitle

\begin{abstract}
We propose a holistic methodology for designing automotive systems that consider security a central concern at every design stage. During the concept design, we model the system architecture and define the security attributes of its components. We perform threat analysis on the system model to identify structural security issues. From that analysis, we derive attack trees that define recipes describing steps to successfully attack the system's assets and propose threat prevention measures. The attack tree allows us to derive a verification and validation (V\&V) plan, which prioritizes the testing effort. In particular, we advocate using learning for testing approaches for the black-box components. It consists of inferring a finite state model of the black-box component from its execution traces. This model can then be used to generate new relevant tests, model check it against requirements, and compare two different implementations of the same protocol. We illustrate the methodology with an automotive infotainment system example. Using the advocated approach, we could also document unexpected and potentially critical behavior in our example systems.      


\end{abstract}
\section{Introduction}\label{sec:intro}
The advent of \emph{connected, cooperative automated mobility} provides a huge opportunity to increase mobility efficiency and road safety.
However, the resulting connectivity creates new attack surfaces that affect the vehicle's safety, security, and integrity.
With an estimated 100 million lines of embedded code, modern vehicles are highly complex systems that need 
to provide consistent cyber-security assurances. Indeed, there are an alarming spike in cyber-attacks targeting connected cars, their electronic control units (ECUs), and the original equipment manufacturer (OEM) back-end servers.

Therefore, making the right security decisions from the early design stages is crucial. The ad-hoc security measures done by domain experts are insufficient to meet the requirements in the automotive domain. 
The standard ISO/SAE 21434 and the mandatory regulation UN R155 advocate for more systematic reasoning about system security.
The United Nations Economic Commission for Europe (UNECE) has adopted new security regulations, such as UNECE R155 and R156, for the homologation of future vehicles that address the identified cyber-attack risks, for example, during software updates.
Similarly, the cyber security standard ISO/SAE~21434, introduced in 2021, defines precise security requirements for vehicles during the entire product life cycle, from its development to its operation and maintenance.
Hence, there is an urgent need for methods and tools that address multiple security-related aspects, from early vehicle design to deployment and operation phases.

This paper proposes a top-down methodology for systematically assessing automotive security at different stages of vehicle development. The proposed methodology follows the product cycle in several steps. 
During the early design phase, we use threat modeling, analysis, and repair to provide more systematic support for the concept design of secure (automotive) systems. These methods allow us to identify the system's weaknesses in security threats and develop structural measures to prevent and mitigate them. 
We then use the threat analysis results to capture the system's critical components concerning security properties and derive a verification and validation (V\&V) plan.
We apply established processes (fuzz testing, penetration testing, etc.) for testing the implemented system components. However, the source code of the component implementation is often unavailable to the V\&V team, and they cannot efficiently use the classical testing methods and tools. In that case, we advocate using automata learning for testing that builds an explainable model of a black-box implementation of a component from a set of executed test cases that facilitates testing and other V\&V activities.
This methodology is a result of a joint research effort amongst
the industrial and academic partners in \TRUSTED\footnote{\url{https://TRUSTED.iaik.tugraz.at/}}, a project focusing on trust and security in autonomous vehicles.
In implementing our proposed methodology, we were also supported by partners from the related LearnTwins\footnote{\url{https://learntwins.ist.tugraz.at/}} project, which focuses on learning-based testing methods for digital twins.


\section{\TRUSTED Methodology}\label{sec:methodology}

\begin{figure}
    \centering
        \includegraphics[scale=0.75]{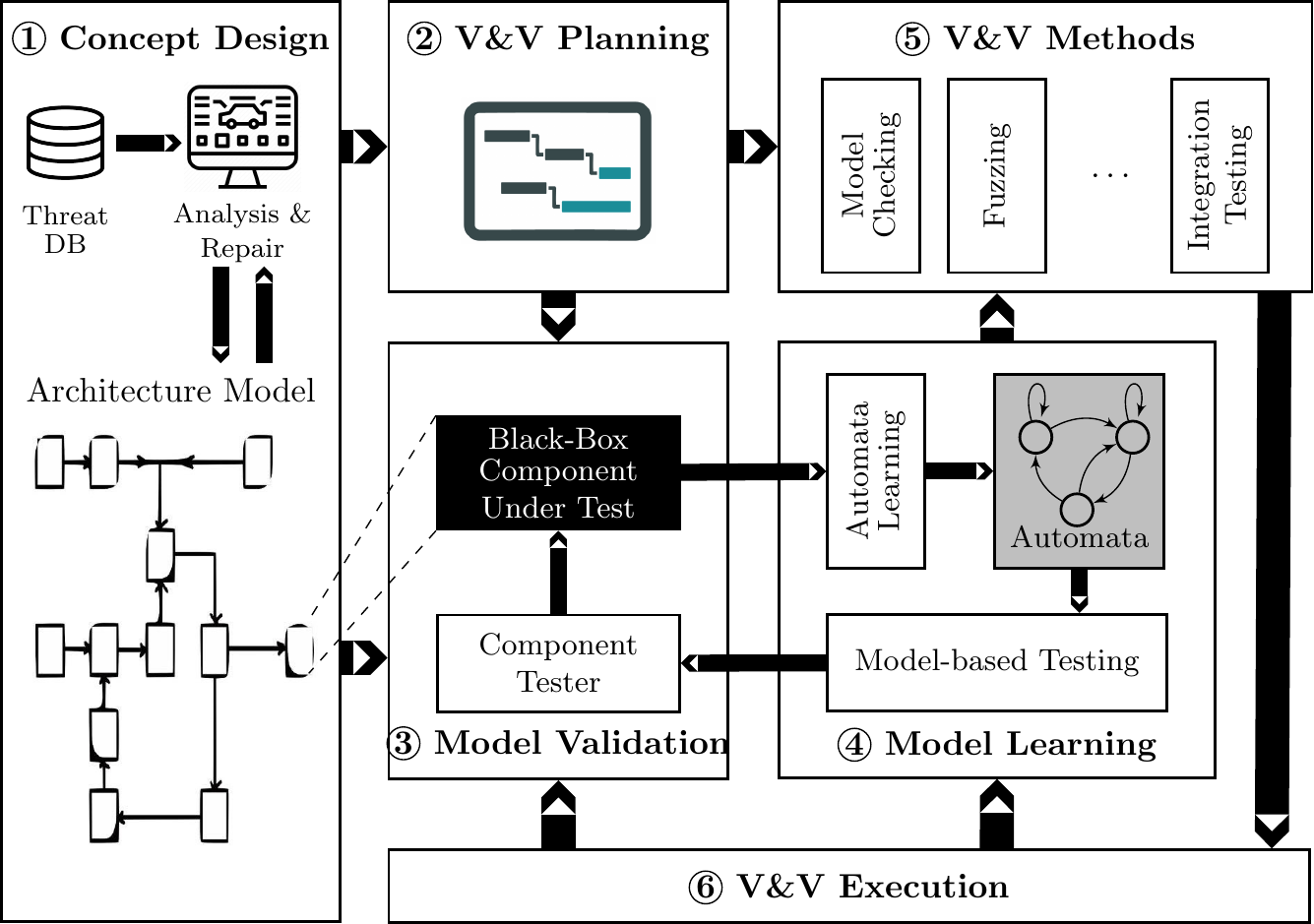}
    \caption{Overview of the \TRUSTED methodology\vspace{-2mm}}\label{fig:overview}
\end{figure}

The \TRUSTED methodology starts with the concept design with a \emph{threat model} of the vehicle; see Stage \circled{1} in Figure~\ref{fig:overview}. 
The threat model consists of two components: (i) a system model architecture and (ii) a threat database.
The system model architecture provides a structural view of the vehicle.
This view includes vehicle components and subsystems (e.g., sensors, actuators, ECUs) and describes their (wireless or wired) interconnections.
We can assign security attributes (e.g., authentication, encryption) to system components and communication links.
A system model can define security boundaries that enclose trusted subsystems and assets we need to protect from potential attacks. 
The threat database contains a set of known threats—these threats from public domain sources, relevant standards, and previous experience. 
The threat model is an input to a threat analysis method allowing the detection of structural weaknesses in the system's architecture. 
We then combine the threat analysis with the repair activities to identify prevention and mitigation actions required to protect the system from identified~threats.

The high-level threat analysis performed in the early stages of the design provides essential insights into the security-related weaknesses in the system architecture. 
We can take structural defense actions to improve the system's security based on threat repair outcomes (e.g., implementing authentication in a specific component).
Yet, there is no guarantee that an attacker cannot break the resulting measures. 
Hence, it is imperative to have a solid verification and validation (V\&V) plan. 
In the \TRUSTED methodology, we use the insights gained by threat analysis and repair to identify risks and prepare an effective V\&V plan corresponding to \circled{2} in Figure~\ref{fig:overview}.


We use the system architecture model developed during the concept design phase to implement and integrate the components of the system. The implementation step is outside the scope of the \TRUSTED methodology, but we assume the components are available as black boxes (see \circled{3} in Figure~\ref{fig:overview}). 
That is, we assume that we can execute components, but we cannot access their implementations.

During the development and integration of different components from the system architecture, verifying and testing safety and security functionalities becomes another critical aspect that we must address.
Model validation (\circled{3} in Figure~\ref{fig:overview}) tests the model for conformance against the component under test.
This step provides either affirmation for the correctness (or completeness, respectively) of the model or counterexamples to refine the latter in a loop until the model is considered good enough to be used for test case generation.

We propose a learning-for-testing approach using automata learning (\circled{4} in Figure~\ref{fig:overview}) as the core method for generating tests during V\&V. 
In automata learning (see Section \ref{sec:learn}), we construct a Finite State Machine (FSM) of the System Under Test (SUT).
We use the inferred FSM to: (1) obtain potential attack data, and (2) 
identify critical inputs that might show differences between the FSM and the SUT.
We must automatically perform the necessary tests during the development and especially the maintenance phase to guarantee a quick response in the event of a threat.

We chose the learning-based testing approach due to its versatility and numerous V\&V activities that we can undertake with the inferred FSM (\circled{5} in Figure~\ref{fig:overview}). 
We can use the inferred FSM to: (1) visualize and understand the implementation, (2) model check it against its formalized requirements (possibly generating test cases on specification violations), (3) generate additional test cases by fuzz testing, and (4) Test for equivalence between implementation and a reference model or another implementation. 

In the last phase (\circled{6} in Figure~\ref{fig:overview}), we use various V\&V strategies to verify the specified properties against the actual component under test. 
The test results are final verification outcomes; meanwhile, we can use them as counterexamples for the learning algorithms in \circled{4} in Figure~\ref{fig:overview}. 
This policy provides a feedback loop for refining the model in the learning-based testing approach.
We execute and store tests using an automated test execution platform that augments generic test cases with additional information.
This additional information comes from a test database or is provided in a grey box testing \cite{marksteinerProcessFacilitateAutomated2021}.

The threat model and the tests created during various design phases must be continuously maintained and updated throughout the vehicle lifecycle. 
We must incorporate new unknown threats and vulnerabilities into the model and re-evaluate the model to find new security issues. 
We must also integrate the changes to functions resulting from software updates into the system model and their impact on the vehicle's security analyzed and re-tested. This closely corresponds with the notions on testing in ISO 21434 and UNECE R155.
\section{Automotive Security by Design}
\label{sec:threat}

In this section, we demonstrate the use of THREATGET~\cite{Schmittner2020}, a tool for threat modeling and analysis to improve the security of automotive applications during their early stages of design (step \circled{1} in Figure~\ref{fig:overview}) and generate an appropriate V\&V plan (step \circled{2} in Figure~\ref{fig:overview}). We illustrate the approach with an automotive infotainment system developed by the industrial partner.

We first model the system using THREATGET  (Section~\ref{sec:arch}) and apply analysis to identify potential structural weaknesses in the system architecture (Section~\ref{sec:threat}). We then use this analysis to derive a V\&V plan (Section~\ref{sec:plan}). Finally, we can augment it with threat repair to propose additional security measures~\cite{repair}.

\subsection{System Architecture Model}
\label{sec:arch}
\begin{figure}[t!]
  \centering
    \resizebox{0.95\linewidth}{!}{\input{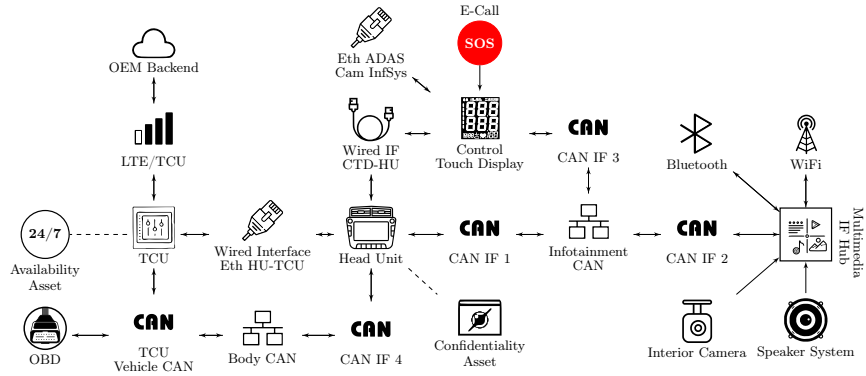}}
 \caption{Automotive infotainment system model.\vspace{0mm}}\label{fig:info}
\end{figure}

We first create an accurate model of the automotive infotainment system (IS),
shown in Figure~\ref{fig:info}.
%
The IS is part of a larger ADAS reference model.
It has several external interfaces that expose an attack surface of the vehicle. The external interfaces in Figure~\ref{fig:info} are Bluetooth, WiFi, Interior Camera, and On-Board Diagnostics (OBD). The Multimedia Interface Hub (MIH) is an essential component of the infotainment system that (co-)implements core functionalities, including navigation, phone calls, and music playback. MIH also bridges external and internal interfaces.
The Telematics Communication Unit (TCU) is the primary interface to the Internet.
Many components in a modern vehicle depend on the TCU.
For example, navigation systems use TCUs to access and update maps, and ECUs use them for over-the-air updates.
Finally, all components except for TCU and Head Unit communicate through a CAN interface.
We add two assets to the model -- the confidentiality asset associated with the Head Unit and the availability asset associated with the TCU. The assets need to be protected, and their associated components are potential targets for attackers.


The IS is a weak security link in modern vehicles because it is more prone to successful cheap attacks than other components (e.g., \textit{Body Control Unit} or the \textit{Engine Control Unit}).
This is due to versatile attack scenarios provided by the use of mainstream Unix-like operating systems, e.g., \textit{Uconnect} and \textit{Automotive Grade Linux}, the user requirements demanding functionalities like a built-in internet browser and installing third-party apps enabling remote code execution attacks, and the use of CAN bus that cannot guarantee communication integrity between the vehicle's external and internal interfaces.


\subsection{Threat Analysis}
\label{sec:threat:analysis}

We analyze the system model with THREATGET against its \emph{threat database}, defining a set of possible threats formulated as \emph{rules}. 
The threat descriptions are collected from multiple sources: automotive security standards and regulations (e.g., ISO/SAE 21434, ETSI, UNECE WP29 R155, and UNECE R156), publicly documented threats identified in past incidents, and expert knowledge.

We illustrate threat rules with two examples used during the analysis of the infotainment system model: the rule named ``Gain Control of Wireless Interface (e.g., WiFi, Bluetooth, or BLE)'' and the rule named ``Flood CAN Communication with Messages''. Both threat rules originate from automotive security analyses performed by domain experts. The first threat's formalization~is
\begin{lstlisting}[
    basicstyle=\ttfamily\footnotesize,
    language=SMT2
]
ELEMENT  : "Wireless Interface"{
    "Authorization" NOT IN ["Yes", "Strong"] & "Input Sanitization" != "Yes" & 
    "Authentication" NOT IN ["Yes", "Strong"] & "Input Validation" != "Yes" &
    PROVIDES CAPABILITY "Control" := "true". }
\end{lstlisting}
This rule specifies that a wireless interface (e.g., WiFi or Bluetooth) that neither implements authorization and authentication nor sanitizes or validates its inputs is susceptible to threats. The last line in the rule explicitly states that if this threat is exploited, the malicious user can control the wireless interface.
The ``Threat Flood CAN Communication with Messages'' threat is formalized as
\begin{lstlisting}[
    basicstyle=\ttfamily\footnotesize,
    language=SMT2
]
FLOW {
    SOURCE ELEMENT  : "ECU" { REQUIRES CAPABILITY "Control" >= "true" } &
    TARGET ELEMENT  : "ECU" {
        HOLDS ASSET {
            "Cybersecurity Attribute" = "Confidentiality" &
            PROVIDES CAPABILITY "Read" := "true" } } &
    INCLUDES ELEMENT  : "BUS Communication" & 
    INCLUDES NO ELEMENT  : "ECU" { "Anomaly Detection" = "Yes". } }
\end{lstlisting}
This rule states that the threat is present if there is a path starting from an ECU that is under the control of a malicious user to another ECU that holds the confidentiality asset and that there is a bus between them and no ECU on the path has implemented anomaly detection.

When applied to the infotainment system model, THREATGET identifies multiple threats. One threat is ``Spoof messages in the vehicle network because of the missing components''. It describes a pattern that starts at an Interface with no Authentication and ends at an ECU with no Input Validation and holds an asset. It includes a wired Shared Medium representing a vehicle's CAN BUS. Moreover, no element (of type Firewall, Server, ECU, or Gateway) on the flow from the Interface to the ECU takes care of Anomaly Detection. 

We can address the identified threats with appropriate security measures. Threat repair~\cite{repair} consists of preventing concrete threats by proposing security measures that can be implemented during the system's design. THREATGET implements \emph{attribute repair}, a method that proposes changes in the components' security attributes as locally deployed measures with a simple cost model. 

In the case of the automotive infotainment system model, e.g., the proposed threat repair measures include enabling authorization and implementing authentication in the WiFi and Bluetooth components.
We note that threat repair does not remove the need for the planned V\&V activities. The fact that authentication is integrated into the WiFi device, following the outcomes of threat repair, does not guarantee that the authentication algorithm's implementation is weakness free. On the contrary, systematic testing of the WiFi's authentication protocol is even more necessary to gain confidence that the WiFi device is not a possible entry point for malicious users.

\subsection{V\&V Planning}
\label{sec:plan}

\begin{figure}[htb!]
    \centering
    \begin{tikzpicture}
[  
    scale=0.72, transform shape,
    every node/.append style = {draw, anchor = west},
    grow via three points={one child at (-0.05,-1.8) and two children at (-0.2,-1.8) and (-2.6,-1.8)},
    edge from parent path={(\tikzparentnode\tikzparentanchor) + (0.4cm,0pt) |- (\tikzchildnode\tikzchildanchor)},
    mirror/.style={parent anchor=south, xshift=-2.4cm},
    parent anchor=south west]
\small

\node [draw, align=left] {\textbf{Confidentiality Asset} \\ Read = true}
      child {node [yshift=0.5cm, xshift=-0.5cm, draw, align=left, fill=gray!30] {\textbf{Head Unit} \\ Control = true}
          child {node [draw, align=left, fill=gray!30, xshift=2cm, yshift=1cm] {\textbf{Head Unit} \\ Updates = yes \\ Managed = no \\ Secure Boot = no \\ Anomaly Detection = no}
              child {node [draw, align=left, xshift=-0.7cm] {\textbf{Multimedia IF Hub} \\ Control = true}
                child {node [align=left, xshift=-0.5cm, xshift=3cm, yshift=1cm] {\textbf{Multimedia IF Hub} \\ Updates = yes \\ Managed = no \\ Secure Boot = no \\ Anomaly Detection = no}
                child {node [draw, align=left, fill=gray!30, xshift=-8mm] {\textbf{Bluetooth} \\ Control = true}
                  child {node [draw, align=left, fill=gray!30,xshift=1.5cm,yshift=1cm] {\textbf{Bluetooth} \\ Input Validation = no \\ Input Sanitization = no \\ Authorization = no \\ Authentication = no}}}
               child [missing] {}
               child {node [draw, align=left, xshift=8mm] {\textbf{WiFi} \\ Control = true}
                  child[mirror, xshift=-2.75cm,yshift=1cm] {node [draw, align=left] {\textbf{WiFi} \\ Input Validation = no \\ Input Sanitization = no \\ Authorization = no \\ Authentication = no}}
                }}
          }}
      }; 
\end{tikzpicture}
    \caption{Attack tree derived from THREATGET. Multiple children from the same node are implicitly interpreted with an OR operation.\vspace{-8mm}}
    \label{fig:at}
\end{figure}
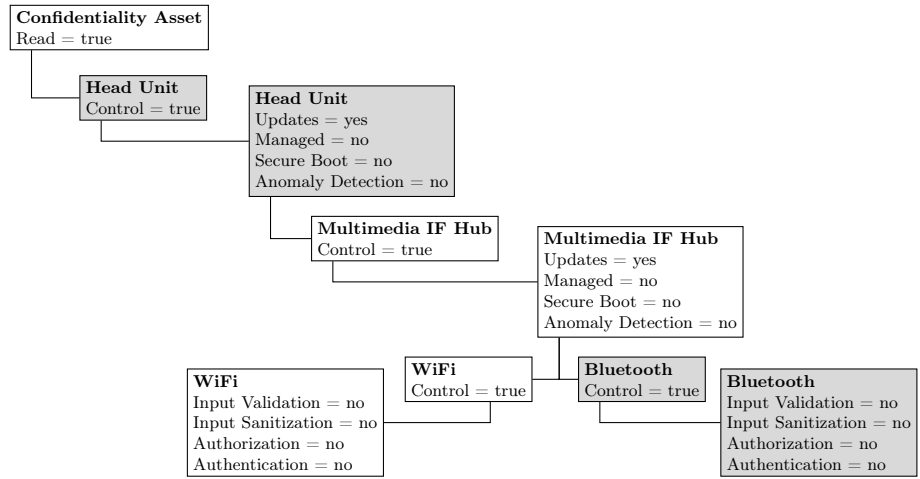

In addition to threat analysis, there is support for identifying and modeling more sophisticated threats using attack trees; c.f. \cite{EbrahimiSCS22}.
This results in more knowledge about potential attackers' steps when intruding into a system.
Simple rules can be assigned attributes called capabilities that are either required for an intrusion or can be gained through the intrusion of a system component.
Moreover, we can define the different access levels to a component (e.g., $Access < Read < Modify < Control$).
Depending on previously acquired capabilities, different attack tree rules trigger, yielding distinct attack trees.
An example of such a generated attack tree is illustrated in Figure~\ref{fig:at}.
%

%

The attack tree depicted in Figure~\ref{fig:at} shows how a malicious user can access the \emph{confidentiality} asset associated with the Head Unit via external interfaces such as WiFi and Bluetooth. For instance, control of the Bluetooth interface can be gained if its security attributes (input validation and sanitization, authorization and authentication) are not implemented or have weaknesses. From there, the user can gain control of the Multimedia Interface Hub, which is not sufficiently secure, and then get control of the Head Unit and hence the access to the asset. The attack tree exposes the most critical components that need to be protected. We note that the attack tree from Figure~\ref{fig:at} is not maximal nor unique -- while THREATGET generates multiple trees for each asset in the model, including the maximal attack trees, we use a simpler tree for illustration purposes.

\section{Automotive Security Testing}
\label{sec:testing}

In this section, we advocate an approach based on learning to test critical components identified by the threat analysis methods during concept design, when these components are assumed to be black-box to the tester. 

\subsection{Automata Learning for Correctness}
\label{sec:learn}

Many cyber-physical components in the automotive domain implement one or multiple finite state machines (FSMs).
%
Implementing larger automotive FSMs becomes cumbersome mainly because: (1) ensuring FSM's correctness w.r.t. its specification is expensive, (2) correctly coding the structure of a large FSM is difficult, and (3) correct integration of FSMs in complex software is hard.
%
Unfortunately, many software-driven components in the automotive industry are black boxes from different manufacturers, hence are hard to verify and thus do not provide functional or non-functional guarantees.

\sm{Should we use formal notations here?}
Given an FSM of a black-box automotive component, we can test and verify it to increase our confidence in its correctness.
Automata learning has proven to be a successful method for learning-based testing of communication protocols that are also used in the automotive domain, e.g., MQTT~\cite{DBLP:conf/icst/TapplerAB17} or Bluetooth Low Energy~\cite{DBLP:conf/fm/PferscherA21}.
%
We use automata learning \cite{angluinLearningRegularSets1987} to infer an FSM model (concretely a Mealy machine)
of the the SUT.
In the learning context we refer to the SUT by system-under-learning (SUL).
In automata learning, a \emph{learner} asks an \emph{oracle} two types of queries.
First, \textit{membership queries} to determine the SUL's output for a given input word.
Second, \textit{equivalence queries} check whether a learned model conforms to the SUL, to which
the oracle returns positive answer or a counterexample. 
A counterexample is an input-output word distinguishing SUL from hypothesis. In practice, oracles for black box systems work with conformance testing.

Ordinarily, real-world systems' alphabets are not manageable for learning algorithms. Abstraction helps to both cope with this fact and to make inferred models more human-readable. Too much abstraction, however, might induce non-deterministic behavior and hide problems we intend to find. There are also automatic abstraction refinement approaches for an optimum of abstraction in a mapper \cite{Aarts:2012,howarAutomataLearningAutomated2011a}.
An abstraction mapper 
consists of a mapping function 
that converts a concrete input 
into an abstract symbol. 
It also observes the SUL's concrete outputs 
and sends an abstraction 
to the learner.
To send a concrete input to the SUL, the mapper inverses the abstraction. 
%
%
There are multiple methods to assess the behavioral correctness of the learned FSMs, including (1) black-box checking \cite{peledBlackBoxChecking1999a}, adaptive model checking~\cite{groceAdaptiveModelChecking2002a}, a combination of learning-based testing and machine learning~ \cite{meinkeLearningBasedTestingCyberPhysical2017a} and symbolic execution~\cite{aichernigAutomataLearningSymbolic2018a}.

\subsection{Use-Case Scenarios}
The attack tree (see Figure~\ref{fig:at}) poses the critical components that need to be tested for security.
In this section, we illustrate our learning-based testing approach on the two components highlighted in gray color in Figure~\ref{fig:at} - the Bluetooth interface (as an entry vector) and the Head Unit ECU. 
\subsubsection{Bluetooth and Bluetooth Low Energy}
\label{sec:BT}
Bluetooth is a well-established standard for wireless audio used in most infotainment systems.
Bluetooth Low Energy (BLE) grows in popularity for car access and sensor data transmission.
%
%
The protocols have a variety of known vulnerabilities \cite{antonioliKNOBBrokenExploiting2019,antonioliBIASBluetoothImpersonation2020,antonioliBLURtoothExploitingCrossTransport2022,seriDangersBluetoothImplementations2017,antonioliKeyNegotiationDowngrade2020}, some also specifically for automotive systems\footnote{\url{https://research.nccgroup.com/2022/05/15/technical-advisory-tesla-ble-phone-as-a-key-passive-entry-vulnerable-to-relay-attacks/}}. 
%
    %

\begin{figure}
    \centering
    \resizebox{0.75\columnwidth}{!}{
    \tikzset{state/.style={draw,rounded corners,rectangle, minimum width=5mm, minimum height=5mm}}
\footnotesize\ttfamily

\begin{tikzpicture} [>=stealth,node distance=1.75cm, on grid]

 
\node (s0) [state, initial] {\adjustbox{max width=0.8cm}{$Established$}};
\node (s4) [state, above=1cm of s0] {$s_3$};
\node (s2) [state, left of= s4] {$s_1$};
\node (s3) [state, below=1.25cm of s0] {$s_2$};
\node (s5) [state, right of=s4] {$s_4$};
\node (s6) [state, below of=s5] {$s_5$};
\node (s9) [state, right of=s5] {$s_6$};
\node (s1) [state, below of= s9] {$s_7$};
\node (s7) [state, right of= s9] {$P_{0}$};
\node (s8) [state, below=1cm of s7] {$P_1$};
\node (s10) [state, below=1cm of s8] {$P_2$};
\node (s11) [state, right of=s10] {$P_{3}$};
\node (s12) [state, right of=s11] {\adjustbox{max width=0.7cm}{$Encrypt$}};
\node (s13) [state, above=1cm of s12] {\adjustbox{max width=0.7cm}{$Pause_{0}$}};
\node (s14) [state, above=1cm of s13] {\adjustbox{max width=0.7cm}{$Pause_1$}};
\node (s15) [state, left of=s14] {\adjustbox{max width=0.7cm}{$Pause_{2}$}};
\node (s16) [state, right of=s8] {\adjustbox{max width=0.65cm}{$Closed$}};
\node (dd) [state, red, below right=0cm and 9.5mm of s6] {$s_{\bot}$};

\path [-stealth, draw, semithick]
    (s0) edge (s3)
    (s0) edge (s2) 
    (s1) edge (s7) 
    (s2) edge (s4)
    (s4) edge (s6)
    (s5) edge (s6)
    (s5) edge (s9)
    (s9) edge (s1) 
    (s6) edge (s9)
    (s7) edge (s8) 
    (s8) edge (s10)
    (s10) edge (s11)
    (s11) edge (s12)
    (s12) edge (s13)
    (s13) edge (s14)
    (s14) edge (s15)
    (s15) edge (s16)
    (s11) edge (s16)
    (s13) edge (s16)
    (s14) edge (s16)
    (s15) edge (s16)
    (s16) edge[loop left, looseness=5] (s16);
\path [-stealth, red, draw, semithick]
    (s6) edge (dd);
    \draw[-stealth, semithick] (s15) -- ++(0,0.5) -- ++(2.5,0) |- (s12.east);
    \draw[-stealth, semithick] (s2.north)++(-0.0625,0) -- ++(0,0.6) -| ($(s7.north)+(-0.125,0)$);
    \draw[-stealth, semithick] (s2.north)++(0.0625,0) -- ++(0,0.3) -| (s5); 
    \draw[-stealth, semithick] (s4.north) -- ++(0,0.9) -| ($(s7.north)+(0.125,0)$);
    \draw[-stealth, semithick] (s3.east)++(0,0.0625) -- ++(0.35,0) |- (s4);
    \draw[-stealth, semithick] (s3.east)++(0,-0.0625) -| (s1);

\end{tikzpicture}
    }
    \caption{Inferred FSM structure for Bluetooth pairing. \vspace{-5mm}}
    \label{fig:bt_model}
\end{figure}
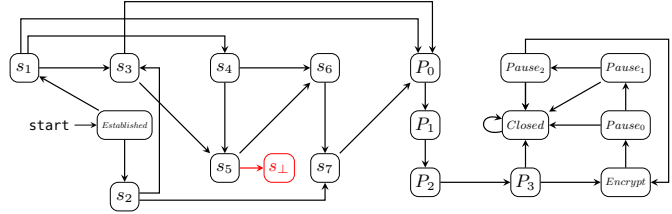

\noindent \emph{Learning Setup}
we use Intel Wireless Controllers (AC 8265 and AX200) implementing  Bluetooth and BLE.
The learning setups are similar, the difference is in the radio hardware and the physical layer, requiring three entities: (1) Radio Device, (2) Learner, and (3) Interface between the two with a mapper. The learner was implemented using the LearnLib framework \cite{isberner_opensource_2015}.

\paragraph{Learned Model and Findings}
We inferred the pairing process models, which are used for encryption and therefore security-critical in the SULs.
As a tangible result,
we discovered a BLE deadlock state (red state in Figure~\ref{fig:bt_model}) in the Linux BLE host software. With repeated out-of-order transmission of pairing requests of different types, we force the respective BLE stack into a state that limits the device to respond to basic link-layer control packets. 
After the state is reached, each following connection will start in this state until the controller is reset. 

\subsubsection{Unified Diagnostic Services}
\label{sec:UDS}

Each ECU has a secure access mode reachable through its UDS implementation, available via vehicle's OBD connector.
%
%
 An attacker able to exploit UDS security features would be also able to manipulate data or even flash the ECU with a malicious firmware.

\paragraph{Learning Setup}
To communicate with the ECU we used a CAN interface.
To learn a different ECU we only need to adapt the interface.
We started by implementing a reduced UDS interface, consisting of instructions to put an ECU into secure access mode. Communications occures via a CAN bus interface. The learner was implemented using the AALpy framework \cite{muskardinAALpyActiveAutomata2022}.

\begin{figure}[b!]
    \centering
        \resizebox{0.75\columnwidth}{!}{\tikzset{state/.style={draw,rounded corners,rectangle, minimum width=5mm, minimum height=5mm}}

\begin{tikzpicture} [scale=1, transform shape, >=stealth,node distance=3.25cm, on grid]
\scriptsize\ttfamily

\node (s0) [state, initial] {$s_0$};
\node (s1) [state, right=1.5cm of s0] {$s_1$};
\node (s2) [state, above=1.5cm of s1] {$s_2$};
\node (s3) [state, right of= s2] {$s_3$};
\node (s4) [state, right of=s1] {$s_4$};
\node (s5) [state, right of=s4] {$s_5$};
\node (s6) [state, right of=s3] {$s_6$};

\path [-stealth, draw, semithick, sloped]
    (s0) edge node[above]{ExtDiag} (s1)
    (s0) edge[loop below] (s0)
    (s1) edge node[above]{Prog} (s2) 
    (s1) edge[loop below] (s1)
    (s2) edge node[above]{SecAcc(\,)} (s3)
    (s2) edge[in=120, out=150, looseness=8] (s2)
    (s3) edge node[below]{\shortstack{SecAcc\\(Key)}} (s4)
    (s3) edge[in=30, out=60, looseness=8] node[above]{SecAcc(\,)} (s3)
    (s4) edge node[above]{Prog} (s2)
    (s4) edge node[above]{SecAcc(\,)} (s5)
    (s4) edge[loop below]  node[left=5pt]{\shortstack{\{~SecAcc(Key),\\\textcolor{red}{SecAcc(Wrong)}~\}}} (s4)
    (s5) edge node[above]{Prog} (s6)
    (s5) edge[loop below] node[left=5pt]{\shortstack{\{~SecAcc(Key),\\\textcolor{red}{SecAcc(Wrong)}~\}}} (s5)
    (s6) edge[red] node[above,red] {SecAcc(PrevKey)}(s4)
    (s6) edge (s3)
    (s6) edge[loop right] (s6);
    
    \draw[-stealth, semithick] (s6.north)++(0,0) -- ++(0,0.9) -| node[above, near start]{\shortstack{\{~SecAcc(NewKey), SecAcc(Wrong)~\}}} (s2); 
    \draw[-stealth, semithick] (s3.north)++(-0.125,0) -- ++(0,0.3) -| node[above, near start]{SecAcc(Wrong)} ($(s2.north)+(0.125,0)$); 

\end{tikzpicture}}
    \caption{Inferred UDS FSM.\vspace{-5mm}}
    \label{fig:uds_model_with_neg}
\end{figure}
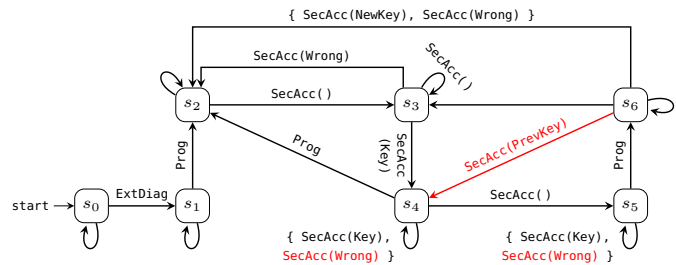

\paragraph{Learned Model and Findings}
The learning experiment resulted in a reduced FSM of the UDS shown in Figure~\ref{fig:uds_model_with_neg}. 
%
An analysis of the results shows that once being successfully authenticated (state $s_4$), an incorrect authentication key will still result in the same state. This is unexpected and allows for prolonging a session without authentication. When requesting a new seed for re-authentication ($s_5$) this behavior persists. Moreover, on re-entering a secure session afterwards (from $s_6$), the ECU accepts an old key as well; an unexpected behavior after re-initiating the key authentication. Figure~\ref{fig:uds_model_with_neg} marks all unexpected behaviors~in~red. 

\section{Conclusion}

We introduced the \TRUSTED methodology for designing and assessing trusted and secure automotive systems. The main novelty of the proposed methodology is its holistic and systematic approach to security, which starts at concept design and is carried down to the implementation and assessment of individual components. We instantiated the different parts of the methodology using the state-of-the-art methods and tools for threat modelling and analysis, automata learning and testing. We illustrated the use of the methodology by applying it step-by-step an automotive infotainment system. Using the learning-based testing approach we could document previously unpublished denial-of-service conditions in the examined BLE setups, as well as unexpected behavior allowing for extending secure UDS programming sessions on the scrutinized ECU.

\paragraph{Future Work}
We plan to further automate the transition from the concept design and V\&V planning on one side, to the actual testing activities done on the level of components by devising a domain-specific test description language that can define abstract V\&V plans derived from the attack trees, and be refined in a way so that eventually it can be executed on a platform (e.g., as in \cite{wolschkeAgnosticDomainSpecific2021a}). Second, the \TRUSTED methodology mainly focuses on the transition from concept design to testing the implementation. We plan to also study the opposite direction -- how to use the component testing results to update the system model and have a more refined threat analysis and a more realistic threat assessment.   
\section*{Acknowledgements}
We thank Andrea Pferscher for her assistance with learning-based testing of the Unified Diagnostic Service protocol.
This research received funding from the program ``ICT of the Future'' of the Austrian Research Promotion Agency (FFG) and the Austrian Ministry for Transport, Innovation and Technology under
grant agreement No. 867558 (project TRUSTED) and within the ECSEL Joint Undertaking (JU) under grant agreement No. 876038 (project InSecTT). The JU receives support from the European Union’s Horizon 2020 research and innovation programme and Austria, Sweden, Spain, Italy, France, Portugal, Ireland, Finland, Slovenia, Poland, Netherlands, Turkey. The document reflects only the author’s view and the Commission is not responsible for any use that may be made of the information it contains.

\bibliographystyle{splncs04}
\bibliography{literature}

\end{document}